\documentclass[aps,nofootinbib,reprint,nobibnotes,showpacs,superscriptaddress]{revtex4-1}

\usepackage[english]{babel}
	\addto\captionsenglish{%
	}

\usepackage{graphicx}

\usepackage{amsmath}
\usepackage{amssymb}
\usepackage[version=3]{mhchem}
\usepackage{mathrsfs}

\usepackage[hidelinks,linktocpage]{hyperref}
\hypersetup{
	colorlinks,
	linkcolor={red!70!black},
	citecolor={green!70!black},
	urlcolor={blue!70!black}
}

\usepackage[usenames,dvipsnames]{xcolor}	
\usepackage{comment}	
\usepackage{upgreek}	

\newcommand{\bb}{$0\nu \beta \beta$} 

\newcommand{\mbb}{\ensuremath{m_{\beta \beta}}} 
\newcommand{\mbbM}{\ensuremath{\mbb^{\max}}} 
\newcommand{\mbbm}{\ensuremath{\mbb^{\min}}} 

\newcommand{\FSigma}{F_{\mbox{\scriptsize \boldmath$\Sigma$}}} 

\newcommand{\el}{e}  

\begin{document}

	\title{Discovery probabilities of Majorana neutrinos based on cosmological data}

	\author{M.~Agostini}
		\email{matteo.agostini@ucl.ac.uk}
		\affiliation{Department of Physics and Astronomy, University College London, Gower Street, London WC1E 6BT, United Kingdom}
		\affiliation{Physik-Department, Technische Universit\"at M\"unchen, 85748 Garching, Germany}
	\author{G.~Benato}
		\email{giovanni.benato@lngs.infn.it}
		\affiliation{INFN, Laboratori Nazionali del Gran Sasso, 67100 Assergi, L'Aquila, Italy}
	\author{S.~Dell'Oro}
		\email{stefano.delloro@unimib.it}
		\affiliation{INFN Sezione di Milano--Bicocca, 20126 Milano, Italy}
		\affiliation{University of Milano--Bicocca, 20126 Milano, Italy}
	\author{S.~Pirro}
		\email{stefano.pirro@lngs.infn.it}
		\affiliation{INFN, Laboratori Nazionali del Gran Sasso, 67100 Assergi, L'Aquila, Italy}
	\author{F.~Vissani}
		\email{francesco.vissani@lngs.infn.it}
		\affiliation{INFN, Laboratori Nazionali del Gran Sasso, 67100 Assergi, L'Aquila, Italy}
		\affiliation{Gran Sasso Science Institute, 67100 L'Aquila, Italy}

	\date{\today}
	
	\begin{abstract}
		We discuss the impact of the cosmological measurements on the predictions of the Majorana mass of the neutrinos, the parameter probed by neutrinoless double-beta decay experiments.
		Using a minimal set of assumptions, we quantify the probabilities of discovering neutrinoless double-beta decay and introduce a new graphical representation that could be of interest
		for the community.
		\\[+9pt]
		Published on: \href{https://journals.aps.org/prd/abstract/10.1103/PhysRevD.103.033008}{Phys.\ Rev.\ D {\bf 103}, 033008 (2021)}
	\end{abstract}

	\maketitle

\section{Introduction}

	The idea that neutrinos are described by a theory that is symmetric between particles and antiparticles, namely the Majorana theory of fermions~\cite{Majorana:1937vz},
	has gained credibility over the years.
	The tiny neutrino masses, proven by the observation of neutrino oscillations, are consistent with the assumption that there is new physics
	at ultrahigh mass-scales~\cite{Minkowski:1977sc,Yanagida:1979as,GellMann:1980vs,Mohapatra:1979ia} and is fully compatible with the gauge structure of the
	Standard Model~\cite{Weinberg:1979sa}.
	A direct way to test Majorana's theory is to search for neutrinoless double-beta decay (\bb)~\cite{Furry:1939qr}.

	Given the importance of \bb, it is of paramount importance to build quantitative projections on the Majorana mass-matrix determining the decay rate.
	In the absence of a conclusive and convincing theory of fermion masses, these predictions should be as much as possible driven by the data and independent by the model.
	This approach was proposed long ago in Ref.~\cite{Vissani:1999tu}, immediately identifying its limitations and initiating their analysis.
	In fact, the Majorana neutrino mass depends on a set of parameters whose uncertainty can strongly limit our capability of making accurate projections.
	These parameters are (i) the mixing angles and mass splittings of neutrino oscillations, (ii) the type of neutrino mass spectrum, (iii) the value of the lightest neutrino mass, and
	(iv) the characteristic phases of the Majorana mass.

	The huge experimental effort of the last decades led to unprecedented accuracies on the measurement of neutrino properties.
	Current-generation \bb\ experiments have reached sensitivities to Majorana masses at the level of $100$\,meV and next-generation searches is foreseen to probe the 
	tens-of-meV region~\cite{Dell'Oro:2016dbc,Snomass:2020}. Precise measurements of the neutrino oscillations have virtually eliminated the uncertainties on the mixing 
	angles and mass splittings relevant for a prediction on the Majorana mass, and global fits are now indicating a mild preference for the normal ordering of the neutrino 
	mass spectrum~\cite{Capozzi:2020qhw,Esteban:2020cvm,deSalas:2020pgw}.
	At the same time, an increasing number of cosmological measurements is converging to constrain the mass of the lightest neutrino.
	
	In this Letter, we revisit and update the discussion on the expectations on the neutrino Majorana mass and their impact on \bb\ searches.
	We propose a new point of view, in which we minimize the amount of assumptions on the unknown parameters -- e.g. the Majorana phases -- which would otherwise require a 
	prior affecting the results (see e.g. Refs.~\cite{Benato:2015via,Caldwell:2017mqu,Agostini:2017jim,DellOro:2019pqi}).
	We set our discussion within the theoretical framework of \bb~mediated by the exchange of ordinary light neutrinos endowed with Majorana masses.
	In fact, the small masses of the three known neutrinos represent today the only unequivocal experimental indication for the existence of physics beyond the Standard Model,
	which is consistent with the overall picture described above. We will also assume that neutrino masses follow the normal ordering.
	The projections for \bb\ assuming inverted ordering are not as interesting because next-generation \bb\ experiments will explore the entire parameter space allowed by this
	model, fully covering any possible projection for \mbb.

\section{Information  from neutrino oscillations}

	The relevant parameter for the \bb~transition is the absolute value of the $\el \el$ entry of the neutrino mass matrix, conventionally indicated with \mbb.
	This parameter can be expressed in terms of the real neutrino masses $m_i$, where $i=1,2,3$, and of the complex mixing matrix elements $U_{\el i}$,
	namely $\mbb = \left| \sum_i U_{\el i}^2 m_i \right|$.
	While the real parts of the matrix elements are constrained by the unitary relation $\sum_i | U_{\el i}^2|=1$, the complex parts are completely unconstrained.
	Given the absolute lack of information on the complex parts, in this work we will base our inferences on the two extreme scenarios in which $\mbb$ assumes the maximum and minimum
	allowed value~\cite{Vissani:1999tu}:
	\begin{subequations}
		\label{eq:mbb_extr}
		\begin{align}
				&\mbbM = \sum_{i=1}^3 \bigl| U_{\el i}^2 \bigr| m_i, \label{eq:mbb_max} \\
				&\mbbm = \max \Bigl\{ 2 \bigl| U_{\el i}^2 \bigr| m_i - \mbbM, 0 \Bigr\} \quad i=1,2,3 . \label{eq:mbb_min}
		\end{align}
	\end{subequations}
	The squared-mass differences $(m_i^2-m_j^2)$ and the parameters $|U_{\el i}^2|$ have been precisely measured by oscillation experiments and can be considered as fixed.
	The uncertainties to these parameters are smaller than $5\%$~\cite{Capozzi:2020qhw}; their propagation would not impact our findings, while it would increase the
	complexity of the computation.
	Thus, Eqs.~\eqref{eq:mbb_max} and \eqref{eq:mbb_min} are a function of only one parameter which is the lightest neutrino mass (i.\,e.\ $m_1$ assuming normal ordering).
	The corresponding {\it true value} of $\mbb$ realized in nature must hence satisfy the condition $\mbbm(m_1) \le \mbb \le \mbbM(m_1)$, where we have made explicit that
	$\mbbm$ and $\mbbM$ are functions depending only on $m_1$.

	\begin{figure}[]
		\centering
		\includegraphics[width=1.\columnwidth]{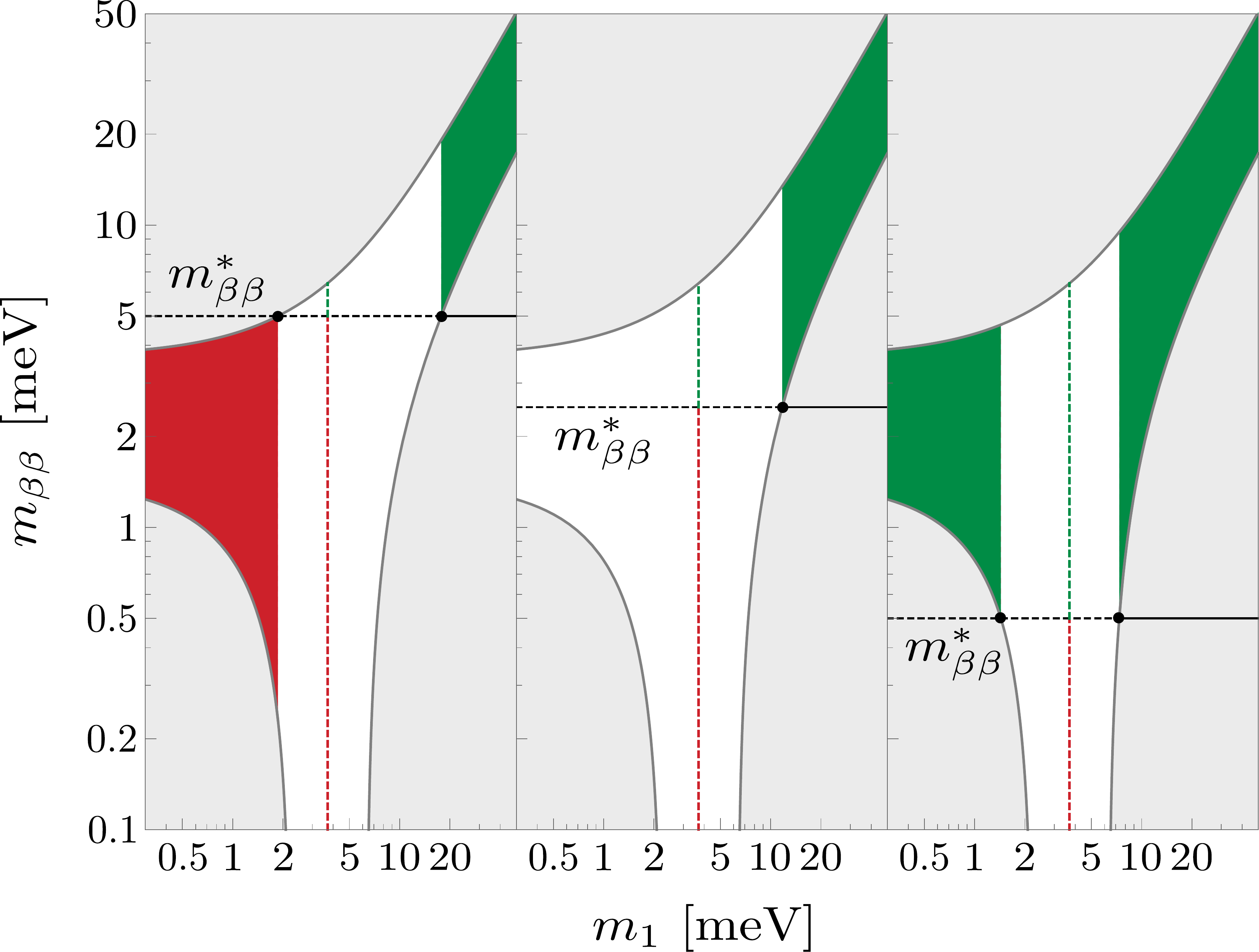}
		\caption{A few representative situations for the search of \bb.
			The illustrative sensitivity values, $\mbb^* = 5, 2.5, 0.5\,$\,meV from left to right, are indicated by the horizontal black dashed lines.
			The black lines enclose the region allowed by neutrino oscillations, assuming normal ordering; the external gray region is forbidden.
			Values above the experiment sensitivity can be probed by the experiment, while those below remain inaccessible.
			Therefore, for each allowed value of $m_1$ and given $\mbb^*$, there are three possible scenarios:
			the experiment will make an observation (green);
			the experiment will not see a signal (red);
			the experimental outcome will depend on the value of the Majorana phases (white).}
		\label{fig:cases}
	\end{figure}

	In the following, we will refer to the smallest \mbb\ value for which an experiment can observe a signal as sensitivity%
	\footnote{The sensitivity is more accurately defined as the median upper limit that can be set at a certain confidence level. The actual definition is however not relevant to the 
		purpose of this work as we will not consider concrete experiments. }
	and indicate it with $\mbb^*$. 
	If we consider a generic experiment that will achieve a sensitivity $\mbb^*$, then the set of possible outcomes of the experiment is limited to three cases:
	\begin{itemize}
		\item $\mbb^* \ge \mbbM(m_1)$: the experiment will not observe a signal even assuming the most favorable value of the Majorana phases; 
		\item $\mbb^* \le \mbbm(m_1)$: the experiment will observe a signal even assuming the most unfavorable value of the Majorana phases;
		\item $\mbbm(m_1) < \mbb^* < \mbbM(m_1)$: the experiment will observe a signal for some values of the Majorana phases but not for others.
			The experiment is exploring an interesting part of the parameter space, however, predictions on whether a signal will be observed or not cannot be done without further assumptions.
	\end{itemize}
	We will further on refer to these cases as {\it inaccessibility}, {\it observation} and {\it exploration}, respectively.
	These scenarios are illustrated in Fig.~\ref{fig:cases} for three illustrative sensitivity values identified in each plot by the black dashed line.
	The red region represents the part of the $(m_1; \mbb)$ parameter space that is not accessible to the experiment; the green one the part for which an observation of \bb~is certain;
	the white one the part for which an observation might (green dash) or might not (red dash) occur.

	Each sensitivity value corresponds to two extreme values for the lightest neutrino mass: $m_1^{\min}(\mbb^*)$ and $m_1^{\max}(\mbb^*)$,
	which are indicated by the black bullets in the figure. Note that $m_1^{\min} = 0$ when $\mbb^*$ lies within the interval $(1.4-3.7)$\,meV.
	The functions $m_1^{\min}(\mbb^*)$ and $m_1^{\max}(\mbb^*)$ can be calculated analytically by solving the fourth-order equation obtained by inverting branch by branch
	Eqs.~\eqref{eq:mbb_max} and \eqref{eq:mbb_min}. However, this procedure does not offer particular advantage, and in this work we solve the equations numerically.
	Uncertainties related to the oscillation parameters are negligible~\cite{DellOro:2019pqi}.

\section{Information  from standard cosmology} 

	The value of the lightest neutrino mass $m_1$ is in one-to-one correspondence with the sum of the neutrino masses probed by cosmology, $\Sigma \equiv m_1 + m_2 + m_3$,
	as the effect of the uncertainties on the mass splittings is negligible.
	The oscillation data provide a lower limit on $\Sigma$ corresponding to the case in which $m_1=0$, i.\,e.\ $\Sigma_{\min} = (58.7 \pm 0.3)$\,meV.
	Therefore, by means of this correspondence, cosmological measurements on  $\Sigma$ can be directly translated in terms of $m_1$, as discussed below
	(see Ref.~\cite{DellOro:2019pqi} for a more detailed discussion).

	The $\Lambda$CDM Model extended with massive neutrinos assumes that neutrinos are stable throughout the life of the Universe and allows to probe the sum of the neutrino masses $\Sigma$,
	provided that we can get cosmological observations both at large scales (CMB) and at the smaller ones (baryon acoustic oscillations, BAO, and Lyman-$\alpha$ forest). 
%
	In several cases, the constraints on $\Sigma$ can be approximated with a Gaussian distribution, in turn corresponding to a parabolic chi-square
	\begin{equation}
	\label{eq:parab_chi2}
		\chi^2 \approx \frac{ ( \Sigma - \overline{\Sigma})^2 }{ (\delta \Sigma)^2 }
	\end{equation}
	where $\overline{\Sigma}$ and $\delta \Sigma$ are known values.
	If we restrict $\Sigma$ to the physical range $\Sigma \ge \Sigma_{\min}$, we obtain the cumulative distribution function (CDF)
	\begin{equation}
	\label{eq:CDF}
		\FSigma (\Sigma) \equiv P( \mbox{\boldmath$\Sigma$} \le \Sigma ) =1 - \frac{ f(\Sigma) }{ f( \Sigma_{\min}) }
	\end{equation}
	where $f(\Sigma)$ is the complementary error function
	\begin{equation}
	\label{eq:erfc}
		f(\Sigma)=\displaystyle \mbox{erfc}\!\left( \frac{\Sigma-  \overline{\Sigma}}{\sqrt{2} \, \delta \Sigma}\right).
	\end{equation}

	The most recent analysis by the Planck Collaboration provides a limit at 95\% confidence interval on $\Sigma$ of $120$\,meV, which has been derived using BAO data and assuming $\Sigma>0$,
	under the approximation that all three neutrinos have the same mass~\cite{Aghanim:2018eyx}.
	An analysis of the same dataset, but with an improved description of the three massive neutrinos imposing the mass differences as fixed by the oscillation experiments,
	gives a bound on $\Sigma$ of 152\,meV at 95\% confidence level (C.\,L.)~\cite{Capozzi:2020qhw}.
	These two results are in good agreement. The results obtained by Planck is in good approximation a Gaussian probability distribution with central value at
        $\Sigma \simeq 0$ and width $\delta\Sigma \simeq 61$\,meV.
	If we now restrict this probability distribution to the physical region $ \Sigma\ge \Sigma_{\min}$, we obtain an upper limit on $\Sigma$ of 146\,meV.
	The $\chi^2$ from Ref.~\cite{Capozzi:2020qhw} is also quasiparabolic and can be approximated -- i.\,e., we get a consistent limit -- by setting [refer to Eq.~\eqref{eq:parab_chi2}]%
	\begin{equation}
	\label{eq:lesin}
		\overline \Sigma = 48.9\,\text{meV}~/~\delta \Sigma = 52.6\, \text{meV}.
	\end{equation}
	Similar results have been obtained already a few years ago by including the  Lyman-$\alpha$ in an analysis similar to those mentioned above.
	The authors of Ref.~\cite{Yeche:2017upn} obtained a limit $\Sigma<140$ meV at 95\% C.\,L.\ by using 
	Lyman-$\alpha$ data from BOSS and XQ-100,
	from an almost-Gaussian likelihood that can be approximated by~\cite{DellOro:2019pqi}
	\begin{equation}
	\label{eq:yechin}
		\overline \Sigma = 41.3\,\text{meV}~/~\delta \Sigma = 49.7\, \text{meV}.
	\end{equation}
	More recently, it has been shown in Ref.~\cite{Palanque-Delabrouille:2019iyz} that the inclusion of the new data from Planck~\cite{Aghanim:2018eyx} could lead to a tighter bound
	of $\Sigma<122$\,meV at 95\% C.\,L.\, (still in the approximation of degenerate neutrino masses).%
	\footnote{Unfortunately, neither the likelihood nor the $\chi^2$ are given in the reference and we cannot include it in our analysis.}
	In the near future, we expect important progress from cosmology, with the possibility of extracting more precise results. These could lead to an uncertainty as small as
	$\delta \Sigma = 12$\,meV \cite{Abazajian:2019eic,Alvarez:2020gvl}. 
	
	The limits extracted by cosmological data appear to be consistent over a large set of assumptions. The use of different hypotheses could make the same bound tighter
	(e.\,g.\ by using a smaller value for $H_0$~\cite{Aghanim:2018eyx,Capozzi:2020qhw,Palanque-Delabrouille:2019iyz}), or looser
	(e.\,g.\ by tuning the intensity of the gravitational wave background~\cite{DiValentino:2015ola}, by allowing a freely varying constant dark energy equation of 
	state~\cite{Gariazzo:2018meg} or by assuming primordial fluctuations to be distributed with a `running spectral index'~\cite{Aghanim:2018eyx,Palanque-Delabrouille:2019iyz}),
	however the most comprehensive results that are currently available lead to  similar conclusions on the sum of the neutrino mass. In particular,
	in both Refs.~\cite{Capozzi:2020qhw} and \cite{Yeche:2017upn}, the value of $\Sigma$ which minimizes the $\chi^2$ is compatible with $\Sigma_{\min}$.
	
\section{Implications for $\mbb$}

	\begin{table}[t]
		\centering
		\caption{Probability of the outcome of an experiment as a function of the terms defined in Eqs.~\eqref{eq:CDF_max} and \eqref{eq:CDF_min} for the three $\mbb$ sensitivity considered
			in Fig.~\ref{fig:cases}.} 
		\begin{ruledtabular}
		\begin{tabular}{l | c c c}
			$\mbb^*$ (Fig.~\ref{fig:cases})		&Inaccess.				&Exploration									&Observation \\[3pt]
			\cline{1-4} \\[-12pt]
			5.0\,meV (left)							&$\FSigma^{\min}$		&$\FSigma^{\max} - \FSigma^{\min}$		&$1 - \FSigma^{\max}$ \\
			2.5\,meV (center)							&$0$						&$\FSigma^{\max}$								&$1 - \FSigma^{\max}$ \\
			0.5\,meV (right)							&$0$						&$\FSigma^{\max} - \FSigma^{\min}$		&$1 - (\FSigma^{\max} - \FSigma^{\min})$
		\end{tabular}
		\end{ruledtabular}
		\label{tab:CDF}
	\end{table}

	We have discussed how \mbb\ and $\Sigma$ are connected to the value of the lightest neutrino mass $m_1$.
	We will now exploit this connection to convert existing information from cosmology into projected discovery probabilities for future \bb\ experiments.
	This can be done by taking a probability distribution on $\Sigma$ and converting it into a probability distribution for \mbb.
	The CDF of \mbb\ can hence be directly interpreted as the probability of observing a signal as a function of the experimental sensitivity.
	
	The connection between $\Sigma$ and \mbb\ is, however, not univocally defined.
	Even considering the oscillation parameters to be perfectly know, we still have no information on the Majorana phases and do not want to introduce unnecessary assumptions on their values.
	For this purpose, we have been focusing on the most extreme scenarios, i.\,e.\ those in which \mbb\ is equal to the maximum or minimum allowed value.
	In these two scenarios, the relationship between $\Sigma$ and \mbb\ becomes univocal and a probability distribution on $\Sigma$ can be converted into one on \mbbM\ or \mbbm.
	Projections for models predicting intermediate values of \mbb\ can be obtained by interpolating between the results for our two reference scenarios.

	To illustrate the procedure in details, let us consider the likelihood described in Eq.~\eqref{eq:parab_chi2} with the parameter values set according to Eq.~\eqref{eq:yechin}%
	\footnote{Incidentally, this allows a direct comparison with our previous results reported in Ref.~\cite{DellOro:2019pqi}.}
	and a particular $\mbb^*$ value.
	We are interested in the two values $m_1^{\min}(\mbb^*)$ and $m_1^{\max}(\mbb^*)$ (black bullets in Fig.~\ref{fig:cases}).
	The discovery probability can hence be calculated using two values of the CDFs in Eq.~\eqref{eq:CDF}:
	\begin{subequations}
		\label{eq:CDF_extr}
		\begin{align}
				\FSigma^{\max} &\equiv \FSigma \left( \Sigma \left( m_1^{\max} \left( \mbb^* \right) \right) \right) \label{eq:CDF_max} \\
				\FSigma^{\min} &\equiv \FSigma \left( \Sigma \left( m_1^{\min} \left( \mbb^* \right) \right) \right). \label{eq:CDF_min}
		\end{align}
	\end{subequations}
	which satisfy the condition $0 \le \FSigma^{\min} < \FSigma^{\max} < 1$ since the CDF is a monotonically nondecreasing function and $0 \le m_1^{\min} < m_1^{\max}$.
	
\begin{figure}[tb]
		\centering
		\includegraphics[width=1.\columnwidth]{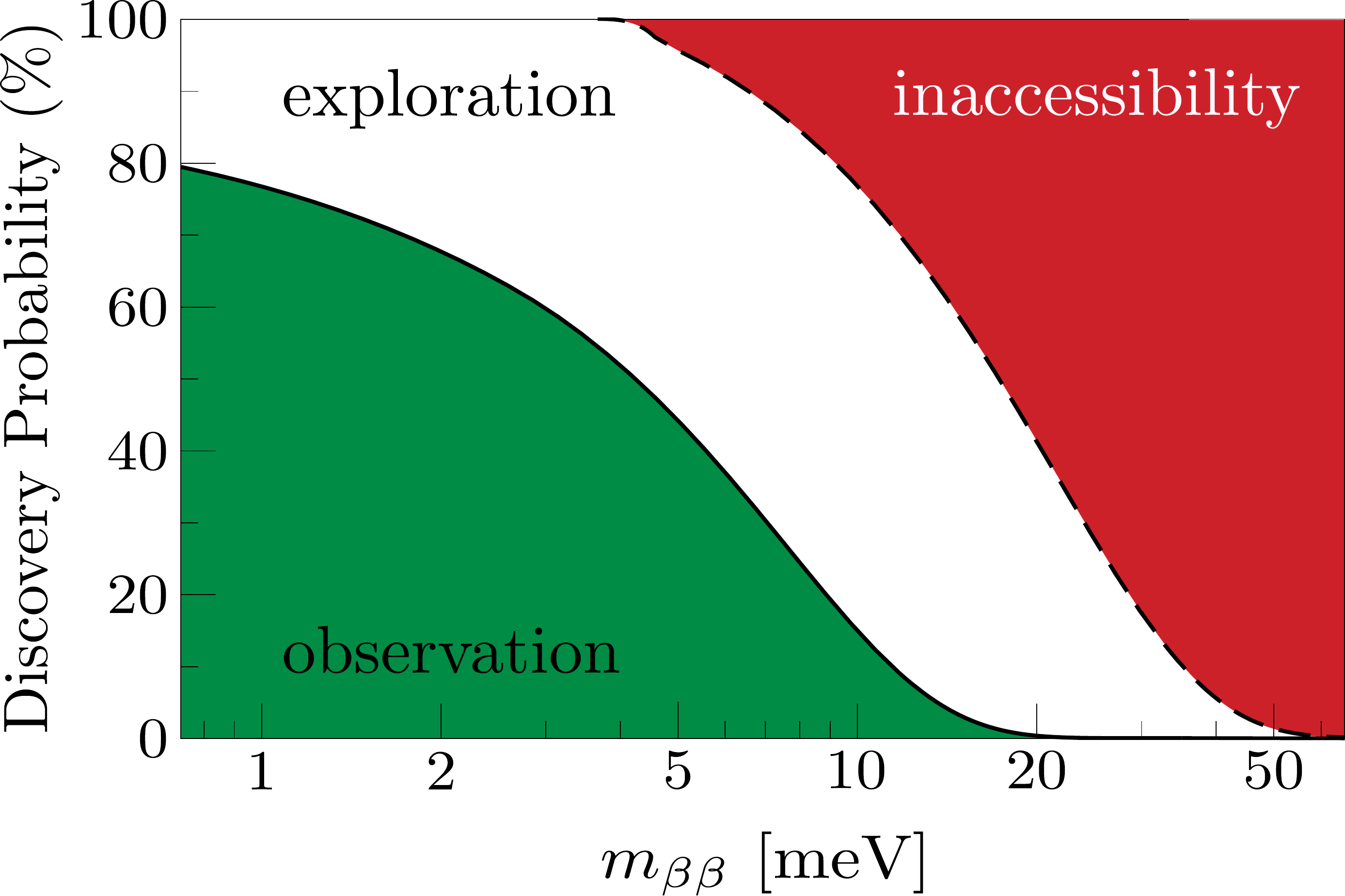}
		\caption{Discovery probability as a function of the experimental sensitivities to \mbb\ for the most unfavorable scenario (black solid line, \mbbm) and the most favorable one
			(black dashed line, \mbbM).
			The colored areas express the probability for the three possible outcomes of an experiment: observing a signal even in the worst case scenario (green, observation),
			not observing a signal even in the best case scenario (red, inaccessibility), and when observing a signal depends on the value of the Majorana phases (white, exploration).}
		\label{fig:flag}
	\end{figure}

	The results are shown in Fig.~\ref{fig:flag} for  $\mbb$ sensitivity spanning from about $1$ to $50$\,meV.
	The solid black line shows the discovery probability for the most unfavorable scenario in which \mbb\ is at its minimal value.
	The black dashed line shows the probability for the most favorable scenario, in which \mbb\ is equal to its maximum allowed value.
	The discovery probability monotonically increases by lowering the \mbb\ value to which an experiment is sensitive.
	Assuming normal ordering, future experiments with sensitivities of the order of 10~\,meV will achieve a discovery power between 20 and 80\%.
	Searches able to reach 5\,meV could reach a discovery power between 50 and 100\%. 
	
	Figure~\ref{fig:flag} can be interpreted also in terms of probabilities of the three possible outcomes of an experiment that have been discussed before: i.\,e.\ 
	{\it inaccessibility} (inaccess.), {\it observation} and {\it exploration}.
	The sum of the probability of these three outcomes is 1 for any value of $\mbb^*$, as the three outcomes are complementary and mutually exclusive.
	Their probabilities can be expressed in terms of the CDF as reported in Table~\ref{tab:CDF}. These probabilities can also be directly read from Fig.~\ref{fig:flag},
	by looking at the width of the red, white and green bands, respectively.
	The smaller the value of $\mbb^*$ is, the larger the green band becomes, signaling an increasing probability of observing a signal even assuming the most unfavorable scenario.
	Conversely, the larger the value  of $\mbb^*$ is, the larger the red band becomes. It should be noted that  even an ``ultimate'' experiment, sensitive to sub-meV values of $\mbb$,
	will have a 20\% probability of being in the {\it exploration} scenario (white band), in which the observation of a signal depends of the value of the Majorana phases.
	This white band does not disappear by decreasing $\mbb^*$ as the true value of $\mbb$ might be negligibly small for some fine-tuned values of the phases and a lightest
	neutrino mass comprised within the range $(2.2-6.3)$\,meV.
	This parameter space would correspond to values for $\Sigma$ in the interval $(61.4-67.5)$\,meV, which is compatible with present cosmological measurements.
	The uncertainty on $\Sigma$ might be reduced to $\delta \Sigma$ of the order of 10\,meV in the near future. However, this will not be enough for a precise measurement of this parameter,
	unless $\Sigma$ will turn out to be large, i.\,e.\ close to the current upper bound from cosmology which would correspond to to \mbb\ values within reach of the future \bb\ experiments.

	The discovery probabilities shown in Fig.~\ref{fig:flag} are also reported numerically in Table~\ref{tab:likelihoods}.
	These values have been computed for a Gaussian probability distribution of $\Sigma$ with the centroid and sigma value of Eq.~\eqref{eq:lesin}.
	The discovery probabilities change by just a few percent when we fix the centroid and sigma to the other values mentioned earlier in this manuscript.
	We have also estimated the impact of the functional form of the probability distribution of $\Sigma$.
	Adopting other reasonable shapes -- e.\,g., a decreasing exponential function -- can change the discovery probabilities by up to 10\% at the \mbb\ value of interest for the next-generation
	experiments, but it does not alter the overall features of our analysis.
	These considerations confirm that our results and conclusions are very robust and valid regardless of which cosmological analysis is used to extract information on $\Sigma$.

\section{Summary}

	\begin{table}[tb]
		\centering
		\caption{Probability (expressed in percent) of the possible outcomes of a search for \bb~given as a function of the experimental sensitivity $\mbb^*$.}
		\begin{ruledtabular}
		\begin{tabular}{r | r r r}
			$\mbb^*$ [meV]				&Inaccess.		&Exploration	&Observation \\[3pt]
			\cline{1-4} \\[-12pt]
			50		&98.7$\,$\%		&1.3$\,$\%		&0.0$\,$\%	\\
			20		&58.6$\,$\%		&41.1$\,$\%		&0.3$\,$\%	\\
			15		&41.9$\,$\%		&55.1$\,$\%		&3.0$\,$\%	\\
			10		&23.1$\,$\%		&62.0$\,$\%		&14.9$\,$\%	\\
			5		&4.4$\,$\%		&51.4$\,$\%		&44.2$\,$\%	\\
			2		&0.0$\,$\%		&32.3$\,$\%		&67.7$\,$\%	\\
			0		&0.0$\,$\%		&12.4$\,$\%		&87.6$\,$\%
		\end{tabular}
		\end{ruledtabular}
		\label{tab:likelihoods}
	\end{table}

	We investigated the impact of cosmological measurements on the possible values of the parameter $\mbb$ and introduced a new procedure (and graphical representation)
	that presents the advantages of minimizing the number of assumptions.
	Our approach is motivated both by the lack of a precise theoretical prediction for the Majorana mass of the neutrino and by the mounting evidence that precision cosmology studies 
	have significant implications on the absolute values of neutrino masses, and hence on the possible values of $\mbb$.
	In the assumption that the exchange of light Majorana neutrinos dominates the rate of the \bb~transition, the results of this work reinforce the view that future-generation experiments
	will have a high discovery power even assuming a normal neutrino mass ordering.
	However, even larger multi-tonne detectors might be needed to find a signal in the less favorable scenarios.

	\begin{acknowledgments}
		We thank Jason Detwiler, Simone Marcocci and Javier Men\'endez for valuable discussions.
		This work has been supported by the Science and Technology Facilities Council (grant number ST/T004169/1), by the Deutsche Forschungsgemeinschaft (SFB No. 1258);
		by the research grant number 2017W4HA7S ``NAT-NET: Neutrino and Astroparticle Theory Network'' under the program PRIN 2017 funded by the Italian Ministero dell'Universit\`a 
		e della Ricerca (MIUR); and by the EU Horizon2020 research and innovation program under the Marie Sklodowska-Curie Grant Agreement No.\ 754496.
	\end{acknowledgments}

	\bibliography{ref_arXiv,ref_Papers,ref_Proceedings}

\end{document}